\documentclass{article}

\usepackage{arxiv}

\usepackage[utf8]{inputenc} 
\usepackage[T1]{fontenc}    
\usepackage{hyperref}       
\usepackage{url}            
\usepackage{booktabs}       
\usepackage{amsfonts}       
\usepackage{nicefrac}       
\usepackage{microtype}      
\usepackage{lipsum}		
\usepackage{graphicx}
\usepackage{natbib}
\usepackage{doi}
\usepackage{amssymb}
\usepackage{url}
\usepackage{booktabs}
\usepackage{multirow}
\usepackage[most]{tcolorbox}
\usepackage{float}
\usepackage[dvipsnames]{xcolor}

\title{Evaluating Prior Art Search with Large Language Models:
Linguistic Complexity and Benchmark Challenges}

\author{
\href{https://orcid.org/0000-0002-6010-670X}{\includegraphics[scale=0.06]{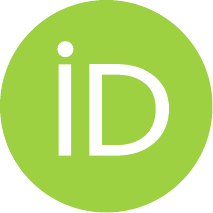}\hspace{1mm}Renukswamy Chikkamath} \\
Hochschule München\\
Lothstraße 34, 80335 München \\
\texttt{renukswamy.chikkamath@hm.edu} 
\And
Linda Andersson \\
Technische Universität Wien\\
1040 Wien, Austria
\AND
Markus Endres \\
Hochschule München
}

\renewcommand{\shorttitle}{EVALUATING PRIOR ART SEARCH WITH LARGE LANGUAGE MODELS}

\hypersetup{
pdftitle={A template for the arxiv style},
pdfsubject={q-bio.NC, q-bio.QM},
pdfauthor={David S.~Hippocampus, Elias D.~Striatum},
pdfkeywords={First keyword, Second keyword, More},
}

\begin{document}
\maketitle
\begin{abstract}

Recent advances in large language models (LLMs) have renewed interest in applying artificial intelligence (AI) to prior art search and other patent information retrieval tasks. While these models demonstrate impressive performance across many natural language processing (NLP) benchmarks, their effectiveness in the patent domain remains closely tied to the design and assumptions of the evaluation datasets on which they are tested. This paper argues that reported performance gains in patent retrieval often reflect benchmark construction choices rather than genuine improvements in retrieval capability or usefulness for patent professionals. We examine this issue through two case studies: the USPTO Explainable AI for Patent Professionals Kaggle challenge and the MoZIP multilingual benchmark for intellectual property. Our investigation of the USPTO challenge shows that extremely high scores can be achieved by exploiting weaknesses in query constraints, token counting, and model-dependent nearest-neighbor ground truth. A critical analysis of the MoZIP benchmark reveals similar concerns, including task formulations that diverge from real-world prior art search, reliance on model-induced relevance definitions, and the absence of strong retrieval baselines. Drawing on prior work in patent information retrieval, particularly the CLEF-IP collections, we emphasize the continued importance of benchmark design principles grounded in realistic workflows, transparent relevance judgments, and reproducible evaluation protocols. The findings highlight the risks of evaluating emerging technologies without adhering to established evaluation standards, even as the field rapidly adopts LLM-based methods.
	
\end{abstract}


\section{Introduction}
\label{sec:Introduction}

An ideal information retrieval (IR) system for patents would accurately capture a user’s information need and return all and only the documents that are relevant, ranked by decreasing usefulness. This goal has long been recognized as aspirational rather than attainable in practice, particularly in complex technical domains. Despite decades of advances in information retrieval research, patent search remains a challenging task, especially for non-expert or infrequent users, due to the volume, complexity, and linguistic characteristics of patent documents \cite{baeza2011ir, manning2008introduction}.

Over the years, successive generations of retrieval technologies have promised to alleviate these difficulties. Early interactive and computer-assisted patent search systems were expected to substantially reduce search effort, yet empirical studies showed that improvements were often incremental rather than transformative. More recently, large language models (LLMs) have generated renewed optimism, driven by their strong performance across a wide range of natural language processing tasks. However, prior IR research suggests that improvements in modeling capacity alone do not necessarily translate into meaningful gains in retrieval effectiveness, particularly when evaluation methodologies are weak or poorly aligned with real-world search practices \cite{voorhees2002philosophy}.

In the patent domain, linguistic barriers pose a persistent challenge for retrieval systems. Prior studies in patent information retrieval emphasize that patent language exhibits substantial variation and ambiguity, even when documents describe similar technical concepts \cite{larkey1999patent, magdy2010examining}. These challenges include:
\begin{enumerate}
  \item \textbf{Synonymy and conceptual variation:} 
  The same technical concept may be described using different terms, paraphrases, or descriptive strategies across patents. 
  For example, an invention disclosed as a ``wireless power transfer system'' may appear in related prior art under expressions such as ``inductive charging,'' ``contactless energy transmission,'' or ``resonant coupling,'' even though these formulations refer to closely related underlying principles. 
  Such variation limits the effectiveness of exact term matching and can result in relevant prior art being missed.

  \item \textbf{Morphological variation:} 
  Patent texts exhibit substantial variation in inflectional, derivational, and compound word forms, particularly in technical domains.
  For instance, a query containing the term ``polymerization catalyst'' should ideally retrieve documents referring to ``polymeric materials,'' ``polymerized compounds,'' or ``catalytic polymer formation.'' 
  Failure to account for these variations can fragment recall and reduce retrieval effectiveness.

  \item \textbf{Anaphora and ellipsis:} 
  Patent descriptions frequently rely on referring expressions and omitted information to avoid repetition, especially in long and complex documents.
  An invention may initially be introduced as ``a heat-resistant composite material,'' while subsequent passages refer only to ``the material'' or ``the composite,'' omitting critical qualifiers.
  When retrieval systems analyze text locally or without discourse-level context, such anaphoric references can obscure the technical meaning of key passages.

  \item \textbf{Lexical ambiguity:} 
  Many terms used in patent documents are polysemous across technical and non-technical domains, leading to ambiguity in retrieval.
  For example, the term ``crown'' may refer to a dental prosthesis, a gear component, a bottle cap, or a monetary unit, depending on context.
  Without effective contextual disambiguation, retrieval systems may return large numbers of irrelevant documents, increasing the effort required to identify truly relevant prior art.
\end{enumerate}

These linguistic phenomena contribute to both low precision and low recall in patent retrieval systems, increasing the effort required to identify relevant prior art. Patent search also differs from general-purpose information retrieval in important structural ways. Patent documents use a specialized and legalistic writing style, and the same invention may be described differently across abstracts, descriptions, and claims. As observed by \cite{larkey1999patent}, inventors may deliberately vary terminology to emphasize novelty or to distinguish their invention from existing prior art, further weakening the relationship between lexical similarity and legal or technical relevance.

While advances in text mining and natural language processing have improved the handling of some of these issues, many systems still struggle to represent the semantic importance of noun phrases and multi-word expressions that are central to patent meaning. Recent research has increasingly focused on LLM-based approaches, often reporting strong performance on newly proposed benchmarks. However, much of this work evaluates models primarily against other neural or LLM-based systems, with limited comparison to strong traditional baselines and insufficient scrutiny of benchmark construction and evaluation protocols \cite{devlin2019bert, reimers2019sentence, thakur2021beir}. As a result, it remains unclear whether reported gains reflect genuine improvements in prior art search or artifacts of dataset design.

This paper argues that meaningful progress in patent information retrieval depends not only on more powerful models, but also on robust and transparent evaluation practices. Through an empirical investigation of the USPTO Explainable AI for Patent Professionals Kaggle challenge and a critical analysis of the MoZIP benchmark, we demonstrate that high performance scores may arise from benchmark artifacts rather than substantive advances in retrieval effectiveness. By situating these findings within the broader context of patent IR research, we emphasize the continued relevance of established benchmark design principles and highlight the risks of evaluating emerging technologies without adequate methodological safeguards.

Taken together, these linguistic and domain-specific challenges underscore that patent retrieval has always been a demanding testbed for information retrieval research, regardless of the underlying technology. While large language models offer new representational capabilities, their evaluation cannot be divorced from the methodological lessons learned through decades of IR experimentation. To assess whether LLM-based approaches constitute a genuine advance for prior art search, it is therefore necessary to revisit the principles, assumptions, and evaluation frameworks established by traditional information retrieval research. The next section provides this foundation by reviewing classical IR methodologies and their role in shaping reliable and reproducible retrieval benchmarks.
\section{Traditional Information Retrieval}
\label{sec:IR}

Traditional information retrieval (IR) research provides the methodological foundation against which modern retrieval systems, including large language model (LLM)-based approaches, should be evaluated. Long before the emergence of neural models, IR developed a well-established experimental paradigm encompassing test collections, relevance judgments, baseline systems, and standardized evaluation metrics. These components were designed to ensure that reported performance improvements reflected genuine advances in retrieval effectiveness rather than artifacts of dataset construction or experimental design.

A central contribution of classical IR research was the formalization of ranked retrieval models and their evaluation under controlled conditions. Models such as the Vector Space Model (VSM) and probabilistic retrieval frameworks introduced principled mechanisms for representing documents and queries, ranking results by estimated relevance rather than relying solely on exact Boolean matching \cite{salton1975vector, robertson1977probabilistic}. These models enabled systematic comparison across approaches and highlighted the importance of strong, transparent baselines. In practice, probabilistic models such as BM25 have proven to be highly robust across domains and remain competitive even when compared to more recent neural approaches \cite{robertson2009probabilistic}.

Equally important was the development of shared evaluation methodologies. Large-scale evaluation campaigns such as TREC established the use of reusable test collections, pooling-based relevance judgments, and metrics such as precision, recall, and mean average precision (MAP) \cite{voorhees2002trec}. These practices made it possible to interpret improvements across systems and over time, while also exposing limitations such as incomplete relevance judgments and topic bias. Subsequent work emphasized that evaluation results must be interpreted in light of collection design, relevance definitions, and task realism \cite{voorhees2002philosophy}.

In the patent domain, these methodological considerations are particularly critical. Prior art search is characterized by high-recall requirements, complex relevance criteria, and substantial linguistic variation. CLEF-IP and related patent retrieval benchmarks explicitly adapted traditional IR evaluation principles to patent-specific tasks, emphasizing realistic search topics, transparent relevance judgments, and comparison against strong lexical baselines \cite{piroi2010clef, magdy2010applying}. Findings from these efforts consistently showed that modest-looking improvements can disappear when task formulations or evaluation protocols change, underscoring the fragility of retrieval gains in this domain.

Despite these lessons, recent work on LLM-based patent retrieval often departs from established IR evaluation practices. New benchmarks may define relevance implicitly through model similarity, restrict query formulations in ways that simplify the task, or omit comparisons with strong traditional baselines such as BM25. As a result, reported performance gains may reflect properties of the benchmark rather than improvements that translate to real-world prior art search. Revisiting traditional IR methodology is therefore essential not to resist new models, but to ensure that their evaluation remains grounded in reproducible, interpretable, and application-relevant standards.

This perspective motivates the analyses presented in the following sections, where we examine contemporary patent retrieval benchmarks through the lens of traditional IR evaluation principles, with particular attention to how benchmark design choices influence observed performance.

\subsection*{\emph{Prior Art Search with CLEF-IP Collections}}

Systematic research on prior art search evaluation predates the recent interest in large language models and has its roots
in large-scale shared evaluation campaigns. Early work on patent retrieval was conducted within the NTCIR framework,
which explored technical paper and patent retrieval tasks and highlighted the challenges posed by long documents,
domain-specific terminology, and recall-oriented search objectives \cite{iwayama2005overview}. These efforts demonstrated that
patent retrieval differs fundamentally from general ad hoc search and requires dedicated evaluation methodologies.

A major European contribution to this line of research was the CLEF-IP track, organized as part of the Cross-Language
Evaluation Forum (CLEF). Introduced in 2009, CLEF-IP focused explicitly on prior art candidate search and provided
large, reusable patent collections, professionally constructed topics, and transparent relevance judgments
\cite{piroi2010clef}. Unlike many later benchmarks, CLEF-IP was designed to reflect realistic examiner workflows,
where high recall is essential and relevance is defined by legal and technical criteria rather than surface-level topical
similarity.
\newpage
Subsequent CLEF-IP tasks extended beyond document retrieval to include passage-level retrieval and text mining
challenges \cite{andersson2013exploring, andersson2016time}. Passage retrieval was motivated by the fact that patent examiners rarely rely on
entire documents alone, but instead assess novelty and inventive step by identifying specific passages that disclose
relevant technical features. These tasks therefore combined elements of ad hoc retrieval and question answering, while
maintaining an explicit focus on professional patent search practice.

Results from CLEF-IP consistently showed that strong lexical baselines, such as BM25-based systems using multiple
document fields and patent metadata, remain highly competitive. In particular, the ``Keep It Simple and
Straightforward'' (KISS) approach demonstrated that substantial recall can be achieved using lightweight indexing
strategies that exploit document structure and classification information, without relying on complex linguistic
processing \cite{magdy2010applying}. More complex systems that incorporated external resources, multiple indices, or
machine learning techniques often achieved higher precision-oriented metrics such as MAP, but not necessarily higher
recall.

These findings reinforce long-standing observations in information retrieval research: shallow linguistic processing,
including stop word removal, stemming, and careful field weighting, can significantly improve retrieval effectiveness,
while deeper linguistic analysis such as part-of-speech tagging or full parsing does not consistently yield gains in
large-scale retrieval tasks \cite{voorhees2002trec, shalaby2019patent, voorhees1999trec}. Importantly, CLEF-IP demonstrated how benchmark
design grounded in realistic tasks and transparent relevance judgments enables meaningful comparison across diverse
retrieval approaches.

Building on these traditional information retrieval principles, the remainder of this paper examines how contemporary
LLM-oriented benchmarks for patent retrieval align or fail to align with established evaluation practices. Section~3
characterises patent documents as a distinct text genre and discusses the implications of their linguistic, structural,
and legal properties for retrieval and relevance assessment. Section~3.1 distills key benchmark design requirements
for evaluating AI and LLM-based retrieval systems in the patent domain. Sections~4 and~5 present detailed case studies
of two prominent evaluation efforts, the MoZIP benchmark and the USPTO Explainable AI for Patent Professionals
Kaggle challenge, highlighting how specific design choices influence reported performance. Finally, Section~6
summarizes the findings and discusses their implications for future benchmark development and responsible evaluation
in patent information retrieval.
\section{The Patent Text Genre}
\label{sec:patenttextgenre}
While traditional IR methodologies provide a general framework for retrieval and evaluation, their application to the
patent domain requires careful consideration of the unique characteristics of patent texts. Patent documents differ
substantially from standard IR collections in both form and function, combining technical disclosure with legal
instrumentation. Understanding these genre-specific properties is essential for interpreting retrieval performance and
for designing benchmarks that meaningfully reflect prior art search tasks.

Patent documents constitute a highly specialized text genre that differs substantially from news articles, web pages, or general scientific literature. These differences have direct implications for information retrieval and text mining methods, particularly in the context of prior art search. A patent is not merely a descriptive document; it is a legal and technical artifact designed to define the boundaries of an invention, often with strategic linguistic choices intended to maximize scope while preserving novelty.

One defining characteristic of patent text is its heavy reliance on complex multi-word expressions and domain-specific terminology. Key technical concepts are rarely expressed through isolated keywords; instead, meaning is distributed across noun phrases, coordinated structures, and carefully constructed claim language. As a result, simplistic lexical representations such as treating words independently in a bag-of-words model often fail to capture the functional semantics of an invention. While such models have shown strong performance in many general-purpose IR benchmarks, prior research has consistently demonstrated their limitations in patent retrieval, where recall of conceptually similar prior art is critical.

Another distinctive feature of patent language is intentional redundancy and variation. The same technical idea may be described multiple times using slightly different wording across the abstract, description, and claims. This variation is not accidental but reflects drafting practices aimed at broad coverage and legal robustness. Consequently, effective patent retrieval systems must account for paraphrasing, synonymy, and structural reformulation at a level that exceeds what is typically required in other text domains.

Patent documents are also long and internally heterogeneous. Claims, abstracts, and descriptions serve different functions and exhibit different linguistic properties. Claims are often highly formalized and syntactically complex, whereas descriptions may include narrative explanations, embodiments, and examples. Treating a patent as a single undifferentiated text field can obscure these distinctions and lead to misleading retrieval behavior, especially when benchmarks do not specify which sections are used for indexing or querying.

Finally, the relevance notion in patent search is fundamentally different from topical relevance as defined in classic IR. Prior art relevance depends on legal and technical criteria, including novelty, inventive step, and claim overlap, which are not always aligned with surface-level semantic similarity. This makes the construction of reliable ground truth particularly challenging and underscores the need for carefully designed benchmarks that reflect professional search practices rather than proxy similarity measures derived from a single model or representation.

These genre-specific characteristics have direct consequences for how relevance is defined and how retrieval systems
should be evaluated. Because prior art relevance depends on legal and technical criteria rather than purely topical
similarity, evaluation benchmarks must be designed with particular care. The following subsection therefore examines
key principles for benchmark design when evaluating AI and large language models in patent retrieval.

\subsection{Benchmark Design for Evaluating AI and LLMs in Patent Retrieval}
\label{sec:benchmark}

Evaluating AI and large language models (LLMs) in patent retrieval requires carefully designed benchmarks that reflect the complexity of real-world patent searches. A robust benchmark should include diverse patent datasets, representative query sets, and evaluation metrics that align with the practical needs of patent professionals. Key elements of such a benchmark include:

\begin{enumerate}
    \item \textbf{Data Collections:} Patent datasets should vary in size and scope, from curated small collections to large-scale repositories like CLEF-IP, covering multiple languages and technical domains. Essential patent sections—titles, abstracts, claims, and descriptions must be included to test different retrieval strategies.
    \item \textbf{Task Categories:} Benchmarks should cover tasks such as Prior Art Search, Patent Passage Retrieval, and validity/invalidity assessment. Each task evaluates an AI or LLM's ability to retrieve relevant documents and identify the specific passages or claims that constitute prior art.
    \item \textbf{Performance Metrics:} Standard IR metrics like precision, recall, and mean average precision (MAP) should be combined with domain-specific measures, such as PRES (Patent Retrieval Evaluation Score). Recall is often emphasized in patent retrieval, as completeness is more critical than precision for establishing novelty.
    \item \textbf{Complex Query Handling:} Benchmarks should test model performance on long, multi-part technical queries that reflect real patent search scenarios.
    \item \textbf{Cross-Domain and Multilingual Evaluation:} Given the diversity of patent domains, benchmarks must assess AI/LLM adaptability across different technical fields and languages.
\end{enumerate}

\subsubsection{MoZIP Benchmark}

\cite{ni2024mozip} introduced the \textbf{MoZIP} benchmark, designed specifically to evaluate LLMs’ understanding of intellectual property (IP) knowledge. MoZIP is multilingual and consists of three complementary datasets:

\begin{enumerate}
    \item \textbf{IPQuiz:} Contains 2,000 multiple-choice questions on IP concepts and regulations, collected from online IP tests in different countries and languages. It evaluates familiarity with IP terminologies, laws, and systems across multiple jurisdictions, covering patents, copyrights, and trademarks.
    \item \textbf{IPQA:} A subset of IPQuiz, comprising 100 frequently asked questions. This set tests LLMs’ ability to respond to practical, real-world IP queries, assessing depth of knowledge and the capacity to interpret regulations and legal scenarios.
    \item \textbf{PatentMatch:} Composed of multilingual patent documents, this dataset assesses whether models can differentiate patents based on their descriptions. It evaluates semantic understanding of technical inventions, which is critical for tasks such as prior art search and patentability assessment.
\end{enumerate}

Together, these datasets provide a rigorous framework to test LLMs’ ability to process complex legal and technical content, reflecting the specialized nature of IP knowledge.

\subsubsection{USPTO Kaggle Challenge Benchmark}

The USPTO’s Explainable AI for Patent Professionals Kaggle Challenge aimed to explore how AI can augment patent retrieval while ensuring results remain interpretable for human examiners. The dataset combined patent metadata (publication number, filing date, family ID, classification codes), patent text (titles, abstracts, claims, descriptions), and precomputed similarity links between each patent and its 50 nearest neighbors (NNs) using embedding models.  

The competition’s objectives were:

\begin{itemize}
    \item Translate AI-generated similarity results into Boolean queries familiar to patent professionals, ensuring explainable and actionable outputs.
    \item Assess AI and LLM capabilities in retrieving specific sets of patents while maintaining relevance and minimizing irrelevant retrievals.
    \item Address challenges in handling large, diverse patent corpora spanning multiple technical fields, languages, and document types.
    \item Evaluate models under limited training data conditions, reflecting realistic scenarios where examiners or LLMs may not have access to extensive labeled examples.
\end{itemize}

This benchmark highlights the trade-offs between automation and interpretability. While embedding-based similarity can identify semantically related patents, mapping these results into Boolean logic is critical for professional usability and trust.

\subsubsection{Benchmarking Considerations}

Designing benchmarks for AI and LLMs in patent retrieval must account for:

\begin{itemize}
    \item \textbf{Interpretability:} Patent professionals require transparent reasoning, which AI models often fail to provide without conversion into Boolean or explainable search formats.
    \item \textbf{Linguistic Complexity:} Patents contain highly technical and domain-specific vocabulary, multi-word expressions, and diverse writing styles.
    \item \textbf{Data Diversity:} Models must perform well across multiple patent families, IPC classes, and languages.
    \item \textbf{Evaluation Metrics:} Both traditional IR metrics and domain-specific measures like PRES must be considered to accurately reflect retrieval quality.
\end{itemize}

By integrating these considerations, benchmarks such as MoZIP and USPTO Kaggle provide complementary frameworks for evaluating LLMs, combining legal and technical knowledge assessment with retrieval performance in complex, real-world scenarios.

Against this backdrop of benchmark design considerations, existing evaluation efforts can be examined more critically.
The MoZIP benchmark represents a recent and influential attempt to assess large language models in the intellectual
property domain. However, when evaluated through the lens of traditional IR methodology and prior art search
requirements, several limitations emerge. The next section analyzes these issues in detail.

\section{MoZIP Benchmark Issues}

The MoZIP benchmark represents an important effort to explore the application of large language models in the intellectual property domain. However, when examined from the perspective of information retrieval and prior art search evaluation, several methodological limitations emerge that complicate the interpretation of reported results. These issues mirror, in part, the challenges observed in the USPTO Explainable AI for Patent Professionals benchmark and highlight broader concerns in the design of recent IP-oriented evaluations.

A primary limitation of MoZIP lies in its task formulation. While the benchmark is positioned as an evaluation of LLMs in intellectual property, its three constituent tasks IPQuiz, IPQA, and PatentMatch, do not correspond to real-world prior art search workflows. IPQuiz and IPQA evaluate domain knowledge and short-form question answering, which, although relevant to general IP education, are fundamentally different from retrieval-driven patent examination tasks. PatentMatch, while closer in spirit to patent similarity assessment, is framed as a multiple-choice classification problem rather than an open-ended retrieval task. As a result, the benchmark favors language understanding and decision-making under constrained options, rather than the recall-oriented, iterative, and exploratory nature of professional prior art search.

The multiple-choice design used across MoZIP introduces a strong structural bias toward LLM-based approaches. By restricting candidate answers to a small, curated set, the benchmark removes many of the challenges that define patent retrieval, such as query formulation, ranking over large collections, and managing lexical and conceptual variation. This design choice makes it difficult to assess whether a model can genuinely retrieve relevant prior art from a corpus, as opposed to selecting the most plausible option from a limited list.

In the PatentMatch task, additional concerns arise from the construction of ground truth relevance. Candidate patents are selected using a combination of BM25 rankings, dense vector similarity based on a proprietary embedding model, and IPC classification overlap. Models are then evaluated on their ability to choose among options that have already been filtered and structured according to these mechanisms. This introduces a form of circularity: relevance is implicitly defined by the behavior of specific retrieval models and classification systems, rather than by independent expert judgments. Similar to the nearest-neighbor dependency observed in the USPTO benchmark, this approach risks evaluating a model’s ability to reproduce the assumptions embedded in the benchmark construction process rather than its effectiveness in identifying meaningful prior art.

The reliance on IPC classifications as a proxy for patent relevance further limits the validity of the evaluation. While IPC codes are useful for administrative organization, they are not reliable indicators of novelty overlap or inventive similarity. Patents within the same IPC subgroup may be legally irrelevant to one another, while highly relevant prior art may fall under different classifications. Established patent retrieval benchmarks, such as CLEF-IP, have avoided treating classification overlap as ground truth for relevance, instead relying on examiner judgments and citation-based relevance assessments.

Another notable limitation is the absence of strong non-LLM baselines. Although BM25 and dense retrieval methods are used internally during dataset construction, they are not evaluated as competing systems within the benchmark. Without reporting standard retrieval baselines or lexical methods adapted to patent search, it is difficult to contextualize claims regarding LLM performance or underperformance. Consequently, observed differences between models may reflect benchmark artifacts rather than substantive advances in retrieval capability.

Finally, the benchmark exhibits inconsistencies in evaluation rigor and scale. The IPQA task relies on a small test set with human judgments but provides limited detail regarding annotation guidelines or agreement. Multilingual coverage is uneven across tasks, and the use of patent sections (e.g., abstracts versus claims or descriptions) is not consistently specified. These factors further complicate reproducibility and cross-method comparison.

Taken together, these issues suggest that while MoZIP is valuable as a domain-specific language understanding benchmark, its suitability as an evaluation framework for patent retrieval and prior art search is limited. As with the USPTO Kaggle challenge, the observed results appear to be strongly influenced by benchmark design choices, underscoring the need for evaluation methodologies that more closely align with real-world patent search practices.

The methodological issues identified in MoZIP are not unique to that benchmark. Similar patterns can be observed in
other contemporary evaluations that aim to assess explainable AI and LLM-based approaches for patent search.
To further illustrate how benchmark design choices affect reported performance, the following section investigates the USPTO Explainable AI for Patent Professionals Kaggle challenge.

\section{Investigating USPTO Kaggle Challenge}
\label{sec:Method}
This section outlines the approach for the USPTO-Kaggle challenge, focusing on its objectives and the performance of top-scoring submissions. We review various methodologies and compare natural language processing (NLP) techniques used for patent search, assessing their effectiveness and supporting the responsible adoption of explainable AI within the intellectual property (IP) field.

The challenge aimed to generate Boolean queries that accurately represent sets of patent documents. Given a query and its 50 nearest neighbors (NN) identified by an AI model, the objective was to create Boolean queries that retrieve the same set of NN patents. The competition's leaderboard showcased high MAP@50 scores, with the top six achieving values between 95.90 and 99.89. These exceptional results not only underscore the efficacy of Boolean queries but also highlight critical research questions regarding the mechanisms by which such queries attain remarkably high retrieval precision. In the following subsections, we outline the data (stage-I), query generation techniques (stage-II), and the results from reproducing the 6th top submission team (sash\footnote{\url{https://www.kaggle.com/competitions/uspto-explainable-ai/leaderboard?}}), along with a comparative analysis of other NLP approaches used in the context of Kaggle competition (stage-III) as depicted in Figure \ref{fig:methodology}. The below subsections also describe Figure \ref{fig:methodology} in greater detail. 
\subsection{Data}
\label{subsec:Data}
For the Kaggle challenge, the USPTO created a NN file with 13,215,512 unique query publication numbers, current through July 2023, where each query has 50 neighbors. All entries in the NN file represent US publication numbers (applications). Additional details, such as metadata and full text, are stored separately, as indicated at the top of Stage I. For evaluation, the USPTO provided a test file of 2,500 samples, a subset of the NN file, with queries published only on or after 1975. However, the complete test file and Whoosh evaluation index were not released to the community.
\begin{figure}
    \centering
    \includegraphics[width=0.8\textwidth]{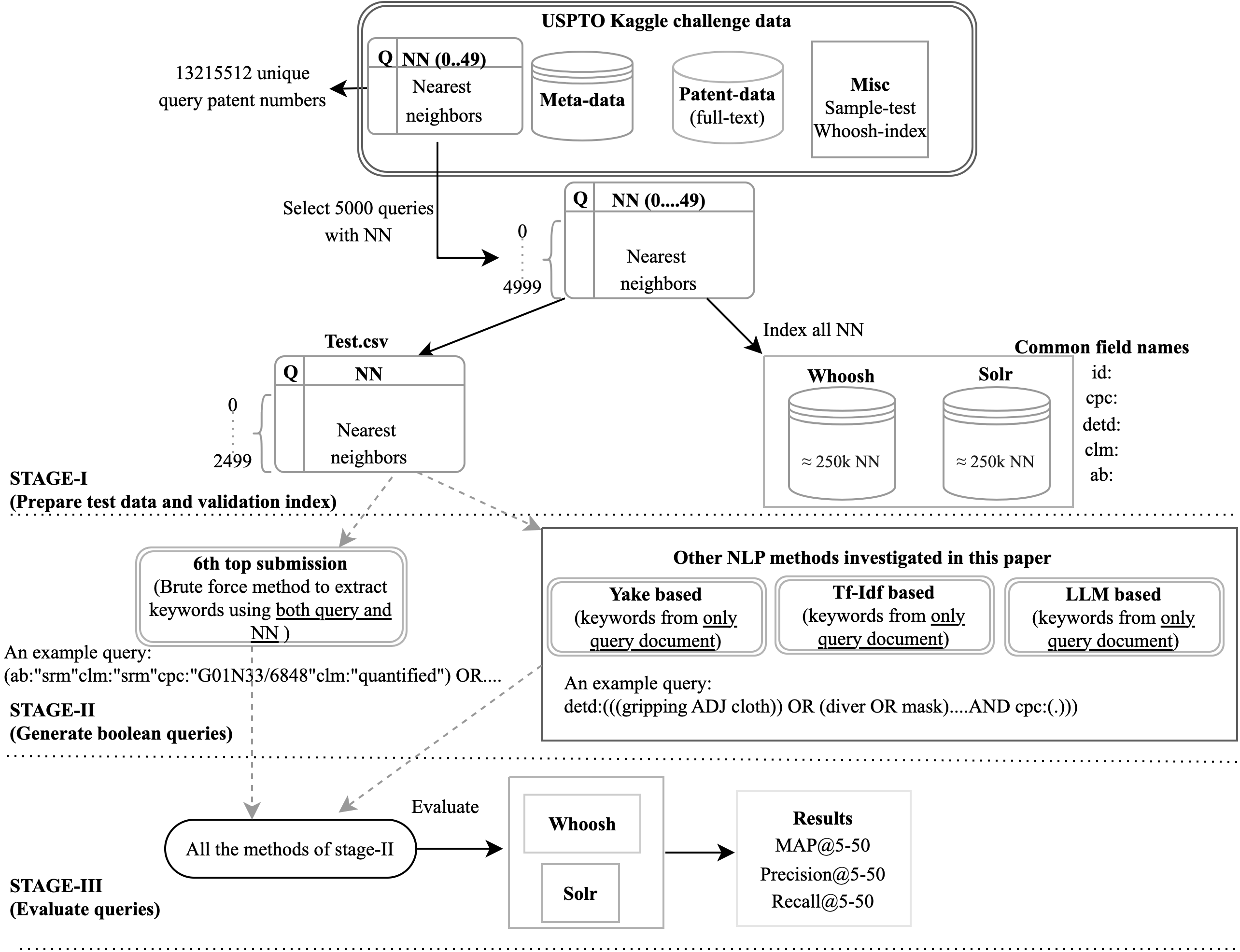}  
    \caption{Detailed description data collection, query generation and evaluation on Whoosh and Solr.}
    \label{fig:methodology}
\end{figure}
To reproduce the experiments, we randomly selected 5,000 samples from the NN file, from which 2,500 samples were further chosen for the test.csv, as shown in the middle of Stage I in Figure \ref{fig:methodology}. To enhance database diversity beyond the neighbors listed in test.csv, we included an additional 2,500 samples for indexing, leading to a total of 5,000 samples indexed in Whoosh. We also indexed these neighbors in Solr 9.6.1, resulting in approximately 250,000 samples (5,000 queries × 50 neighbors) in both indices. Initially, the challenge tested participant queries solely on the Whoosh index; however, to analyze query behavior across different databases, we indexed in Solr using the same column fields (id, cpc, detd, clm, ab), as shown in the right-middle of Stage I in Figure \ref{fig:methodology}. This test data was used to generate queries and evaluate them on both Whoosh and Solr indices, as detailed below.

\subsection{Query Generation}
\label{subsec:QueryGeneration}

Query generation consists of two broad steps: selecting terms, tokens, or words, and formulating a query using those terms. To facilitate this process, the USPTO established the following conditions: i) Only supported search operators are Boolean (OR, AND, NOT, XOR), proximity (ADJ, NEAR), and wildcards (+, ?, \$); ii) The maximum number of tokens in a query is 50, in accordance with the Whoosh function (\emph{whoosh\_utils.count\_query\_tokens)}; iii) Each query must retrieve 50 nearest neighbors; if there are fewer than 50, results are padded with non-matches to reach 50, or truncated if there are more.

We employed two primary methods to generate queries: one involves using solutions from the winning teams in the Kaggle competition, while the other consists of NLP-based experiments aimed at comparative analysis of this challenge across different settings and scenarios.

\subsubsection{Winning Team Solutions}
\label{subsubsec:Winning team solutions}
Following the challenge, we conducted a detailed investigation of the top-scoring teams (the top six) to understand their query generation methods. The methodologies that achieved scores such as 99.89 MAP@50 are particularly intriguing and warrant further exploration. The challenge itself is complex, constrained by a limit of 50 tokens per query approximately half of which are typically operators while also necessitating the identification of the nearest 50 neighbors. This limitation makes it challenging to select representative words for those 50 neighbors, significantly increasing the likelihood of including irrelevant terms. Our investigation yielded several key observations:
\begin{enumerate}
    \item \textbf{Exploitation of Token Counting Flaws:} For instance, consider the query from the highest-scoring team\footnote{\url{https://www.kaggle.com/competitions/uspto-explainable-ai/discussion/522233}} (Team TK and T0m, with 99.89 MAP250), ``ab:spectrometry AND ti:mass AND detd:cid AND detd:quadrupole AND clm:fragments AND detd:analyzers AND detd:qu -antitation AND detd: isotope AND clm:fragment''. This query counts as 17 tokens according to the standard Whoosh function, which relies on whitespace for tokenization. However, participants circumvented this limitation by using special characters (e.g. '-') in place of whitespace, effectively reducing the token count. For example, the same query could be reformatted as ``ab:spectrometry-ti:mass-detd:cid-detd:quadrupole-clm:fragments-detd:analyzers-detd:quantitation-detd:isotope-clm:frag -ment'', which counts as a single token. Such flaws and weak benchmark settings undermines scientific rigor and transparency, allowing participants to bypass the challenge's established conditions.
    \item \textbf{Reliance on Brute Force Methods:} The top-scoring solutions not only utilized the token counting flaw but also predominantly employed brute force techniques to exploit the contents of both queries and their neighbors. Given that the USPTO disclosed the entire NN file along with mappings of queries to their neighbors, participants were able to develop methodologies that exploited predefined queries and results. This allowed for the creation of a statistical database identifying unique lexical features associated with each patent. For example, a micro-query from the highest-scoring team, \emph{detd:lumos-detd:lysc-cpc:G16B40/10}, demonstrates a unique word combination that, when executed in Whoosh, retrieves only US-2021239708-A1. The exhaustive nature of this brute force lexical analysis is evident, as “lumos” appears only once in the entire patent, and rest is unique combination.
\item \textbf{Evaluation Focus on Irrelevance:} Based on initial evaluations prior to final submission, participants noted that the challenge's scoring was heavily influenced by the retrieval of zero irrelevant results. For instance, the second-place team's (Shun\_PI\footnote{\url{https://www.kaggle.com/competitions/uspto-explainable-ai/discussion/522258}}) performance indicated that achieving 25 correct targets alongside 25 incorrect ones resulted in a MAP@50 of only 50\%, while 25 correct targets with zero incorrect targets yielded a MAP@50 of 84\%. This finding led participants to exploit the complete database to avoid irrelevant results, as previously mentioned. Consequently, they adopted a micro-query-based approach for query formulation, structured as \emph{(micro-query1) OR (micro-query2) OR (...micro-queryN)}.

\item  \textbf{Impact of Micro-Query Strictness:} Typically, the stricter the micro-query, the higher the likelihood that the retrieved result will be singular. By using the OR operator, participants could generate around 25 micro-queries for a single query, ideally retrieving 25-30 correct targets. Coupled with the observation regarding the exclusion of irrelevant results, achieving 25-30 correct targets with zero incorrect targets would enable the USPTO to append non-matches, as stipulated in the challenge’s third condition. However, this raises an unresolved question: if micro-queries yield only a single result, how does the USPTO calculate scores when the padding condition is considered, allowing for scores as high as 99\%?
\end{enumerate}
   
\subsubsection{Other NLP Methods}
\label{subsubsec:Other NLP Methods}
Based on the observations made regarding Kaggle solutions for the USPTO challenge, we aimed to adhere strictly to good scientific practices in the context of AI-based patent search settings. Our investigation focused solely on the content of query patents, deliberately avoiding any flaws inherent in the databases. We sought to identify meaningful words or tokens from query patents without resorting to brute force statistical preprocessing that relies on pre-known answers. To achieve this, we experimented with open-source libraries such as YAKE\footnote{\url{https://github.com/LIAAD/yake}}, employed a straightforward TF-IDF method, and utilized large language models (LLMs) to generate important keywords from patent texts. Notably, we discovered that some patents predating 1975, such as US-1081085-A, contain no text beyond the disclosure in their descriptions. Consequently, we chose to focus on the descriptions of patents lacking claims, which led us to employ “detd:(...)” styled queries.

In our exploration of various NLP practices, we adhered to a general query generation method that involved extracting keywords and formulating queries. The extraction of keywords was conducted using several approaches, detailed below:
\begin{enumerate}
    \item \textbf{YAKE:} It is an unsupervised keyword extraction method that is corpus-independent. It has outperformed several unsupervised approaches (e.g., TF-IDF, KP-Miner, RAKE, TextRank) as well as supervised methods \cite{campos2020yake}. Therefore, we applied this method to patent claims and descriptions to identify 25 keywords, using the following experimental settings: language = "en", max\_ngram\_size = 2, deduplication\_threshold = 0.9, deduplication\_algorithm = 'seqm', windowSize = 1, and numOfKeywords = 25. We also investigated the role of stopword removal on retrieval performance when searching using Boolean queries. Using YAKE, we produced a set of keywords, including unigrams and bigrams, which were subsequently utilized in query formulation.
    \item \textbf{TF-IDF:} We utilized the Solr index to calculate the TF-IDF values for the patent queries, selecting the top 25 words with the highest TF-IDF scores for further use in query formulation. To evaluate TF-IDF, we employed the Solr function “score” which is its default ranking algorithm (usually a version of TF-IDF or BM25). This score is computed based on several factors, including term frequency (TF), inverse document frequency (IDF), and additional factors such as proximity, boost factors, and query complexity. We selected those words with the highest scores.
    \item \textbf{Large Language Models (LLMs):} To leverage the advanced language capabilities of LLMs, we used the Mistral 7B instruction-based model to extract important keywords from patent texts. We also explored using LLMs to prepare queries directly. However, we encountered several irregularities in the structure of the required queries for Whoosh-like databases, despite employing various prompt engineering practices. The challenges included misplaced brackets and the inclusion of unnecessary operators, adding complexity to verifying the competence of queries for our experiments with Whoosh and Solr databases. Consequently, we limited the use of LLMs to the extraction of important keywords that could retrieve nearest neighbors or similar patents. 
\end{enumerate}

Once the keywords are extracted, query formulation requires some preprocessing, such as using the ADJ operator for every bigram. For example, the bigram ``slot card” will be translated to ``slot card''\textasciitilde5 for the Solr database and (slot ADJ5 card) for Whoosh. For unigrams, we used the OR operator in between. We also conducted experiments using CPC information in the queries. Ultimately, the query must adhere to a token limitation of 50. The prompt we used (refer below textbox), a sample formulation and example query are shown below:
\begin{tcolorbox}[colback=gray!5!white,colframe=white!75!black]

prompt = f``````
    Task:
    
    Imagine you are patent professional with linguistic experience. Analyze and understand the given patent claims. Extract important keywords (either unigrams and bigrams) that should yield this patent when searched in the United States Patent and Trademark Office (USPTO) database. Also generate similar keywords to those extracted keywords, such that similar patents can be retrieved.

    Instructions:
    
    1. Carefully read and understand the provided claims.
    
    2. Identify the key terms, concepts, and components that are likely to be used when searching for this or similar patents.
    
    3. After the `Keywords' section below, write only the keywords, each separated with a comma (`,'). For example, `keyword1, keyword2, keyword3\_1 keyword3\_2'.
    
    4. You are not allowed to add any narratives or text before or after your response.

    * claims: {row.claims}

    Keywords:
    """

\end{tcolorbox}
Query formulation: \emph{detd:((bi-grams) AND (uni-grams) AND (CPC))}

Query example: \emph{detd:(((tartaric ADJ acid) OR (nitrilopropionic ADJ acid) OR (ethylenediaminetetraacetic ADJ acid) OR (citric ADJ acid) OR (chelating ADJ agent) OR (complexing ADJ agent) OR (pharmaceutically ADJ acceptable) OR (N-2-hydroxyethyliminodiacetic ADJ acid) OR (radioactive ADJ erbium-165) OR (serum ADJ albumin))) AND cpc:(A61K2123/00 OR A61K51/04 OR G21G1/06 OR G21H5/02)}

\subsection{Evaluation}
\label{subsec:QueryGeneration}
In investigating the nature of the top-scored queries and their approaches, we developed the hypothesis that these solutions exhibit volatility concerning platform dependencies, such as Whoosh. This volatility confines them to the limitations of Whoosh. Additionally, we explored how straightforward methods, which do not exploit the complete dataset (including questions, results, and database) in a brute-force statistical manner, perform in terms of retrieval precision. To this end, we evaluated all methods, including one Kaggle solution (the sixth top solution) and three NLP practice-based approaches, on both Whoosh and Solr. Although the competition primarily focused on Whoosh, conducting evaluations on Solr was necessary for a comprehensive analysis and to assess the real-time application of queries.

The evaluation results for the state-of-the-art (SOTA) method are presented in Table \ref{tab:sota-evaluation}, where the same set of queries and the same validation index were tested on both Whoosh and Solr. This setup remained consistent across all other experiments conducted for a fair comparison in this paper.

\begin{table}[h]
\centering
\caption{Evaluation Results of Brute Forced SOTA method(in percentages), \textbf{text content of both query and its relevant results (50 neighbors)} are used for query generation }
\begin{tabular}{llcccccc}
\toprule
\textbf{Method}   & \textbf{Metric}   & \textbf{@5} & \textbf{@10} & \textbf{@20} & \textbf{@30} & \textbf{@40} & \textbf{@50} \\
\midrule
\multirow{3}{*}{SOTA\_Whoosh} 
& MAP & 99.86 & 99.87 & 99.83 & 99.07 & 88.31 & 72.70 \\ 
& Precision & 99.91 & 99.92 & 99.90 & 99.14 & 88.38 & 72.76 \\ 
& Recall  &  9.99 & 19.98 & 39.96 & 59.48 & 70.70 & 72.76  
\\\cline{2-8} 
\multirow{3}{*}{SOTA\_Solr} 
& MAP & 82.56 & 75.17 & 37.58 & 25.05 & 18.79 & 15.03 \\ 
& Precision & 86.66 & 81.41 & 40.70 & 27.13 & 20.35 & 16.28 \\ 
& Recall  &  8.66 & 16.28 & 16.28 & 16.28 & 16.28 & 16.28  \\

\bottomrule
\end{tabular}

\label{tab:sota-evaluation}
\end{table}
In reproducing the Kaggle experiments with settings similar to those of the USPTO challenge, we observed a variation in Mean Average Precision (MAP). In addition to MAP@50, as specified in the USPTO Kaggle challenge, we also assessed the methods for precision and recall at 50. The SOTA method achieved 94.90\% MAP@50 in the Kaggle competition (ranking 6th on the leaderboard), while reproduced experiments yielded 72.70\% MAP@50. We attribute this variation to the fourth observation hypothesized in Section 3.2.1. There are still ongoing discussions\footnote{\url{https://www.kaggle.com/competitions/uspto-explainable-ai/discussion}} on Kaggle regarding the correctness of the evaluation scheme in the USPTO Kaggle competition.

Precision, in relation to MAP, did not show significant differences between Solr and Whoosh. Whoosh exhibited a gradual increase in recall as the number of top values increased from 5 to 50. However, with Solr, recall plateaued after the top 10 values. From Table \ref{tab:sota-evaluation}, we also observe that the drop in MAP from @5 to @50 (99.86\% to 72.70\%) in Whoosh is 27.16\%, whereas in Solr, the drop is 67.56\%, showing a margin variation of 40.40\%. Table \ref{tab:yake-evaluation} onwards we presents the results of other NLP models we tested for this challenge, using the same experimental conditions and settings as the SOTA evaluation. However, these other NLP methods rely solely on the query text content.
\begin{table}
\centering
\caption{Evaluation Results of YAKE\_Whoosh (\textbf{YW}) and YAKE\_Solr (\textbf{YS}) (in percentages), \textbf{only query text} is used for query generation. \\
YW1: No stopwords removal, ADJ=5, AND used between bi-grams and uni-grams \\
YW2: Stopwords removal, ADJ=1, AND used between bi-grams and uni-grams \\
YW3: Stopwords removal with CPC, ADJ=1, OR used between bi-grams and uni-grams
\\
YS1: No stopwords removal, proximity=5, AND used between bi-grams and uni-grams \\
YS2: Stopwords removal, proximity=1, AND used between bi-grams and uni-grams \\
YS3: Stopwords removal with CPC, proximity=5, OR used between bi-grams and uni-grams}
\begin{tabular}{llcccccc}
\toprule
\textbf{Method}   & \textbf{Metric}   & \textbf{@5} & \textbf{@10} & \textbf{@20} & \textbf{@30} & \textbf{@40} & \textbf{@50} \\
\midrule
\multirow{3}{*}{YW1} 
& MAP & 5.53 & 4.25 & 3.19 & 2.61 & 2.26 & 2.02 \\ 
    & Precision & 8.10 & 7.68 & 7.12 & 6.64 & 6.30 & 6.05 \\ 
    & Recall  & 0.81 & 1.53 & 2.84 & 3.98 & 5.04 & 6.05 \\\cline{2-8} 
\multirow{3}{*}{YW2} 
    & MAP & 7.29 & 5.54 & 4.12 & 3.39 & 2.93 & 2.60 \\ 
    & Precision & 10.28 & 9.38 & 8.62 & 7.99 & 7.50 & 7.07 \\ 
    & Recall  & 1.02 & 1.87 & 3.44 & 4.79 & 6.01 & 7.07 
    \\\cline{2-8} 
\multirow{3}{*}{YW3} 
    & MAP & 36.39 & 29.78 & 22.35 & 17.77 & 14.55 & \colorbox{lightgray}{\textbf{12.24}} \\ 
    & Precision & 43.99 & 39.12 & 31.79 & 26.57 & 22.52 & \colorbox{lightgray}{\textbf{19.47}} \\ 
    & Recall  & 4.39 & 7.82 & 12.71 & 15.94 & 18.01 & \colorbox{lightgray}{\textbf{19.47}} \\
\bottomrule

\multirow{3}{*}{YS1} 
& MAP & 45.96 & 36.57 & 18.28 & 12.19 & 9.14 & 7.31 \\ 
    & Precision & 51.36 & 43.46 & 21.73 & 14.48 & 10.86 & 8.69 \\ 
    & Recall  & 5.13 & 8.69 & 8.69 & 8.69 & 8.69 & 8.69 \\\cline{2-8} 
\multirow{3}{*}{YS2} 
    & MAP & 45.79 & 36.21 & 18.12 & 12.08 & 9.06 & 7.25 \\ 
    & Precision & 50.96 & 43.03 & 21.51 & 14.34 & 10.75 & 8.60 \\ 
    & Recall  & 5.09 & 8.60 & 8.60 & 8.60 & 8.60 & 8.60 
    \\\cline{2-8} 
\multirow{3}{*}{YS3} 
    & MAP & 54.59 & 45.09 & 22.54 & 15.03 & 11.27 & \textbf{9.01} \\ 
    & Precision & 60.69 & 52.98 & 26.49 & 17.66 & 13.24 & \textbf{10.59} \\ 
    & Recall  & 6.06 & 10.59 & 10.59 & 10.59 & 10.59 & \textbf{10.59} \\
\bottomrule
\end{tabular}

\label{tab:yake-evaluation}
\end{table}
From Table \ref{tab:yake-evaluation}, which presents the results for the YAKE method, it is evident that the removal of stopwords improves performance across all metrics (MAP, precision, and recall). Increasing the value of ADJ in Whoosh does not lead to better scores; however, the opposite effect is observed in Solr. Thus, the best results in Whoosh are achieved with a lower ADJ value, whereas in Solr, using a proximity value of 5 yields better scores. Bold numbers indicate the best results within each system (Whoosh and Solr), and the best scores overall between Whoosh and Solr for the YAKE method are highlighted. There is a notable relevance of CPC values in both Whoosh and Solr databases, indicating that CPC information is a valuable feature in patent domain search, contributing to improvements in both precision and recall. In Solr, Recall generally gets saturated at top 10 where as in Whoosh this is not the case. 

Additionally, from Table \ref{tab:yake-evaluation}, we observe a MAP drop from @5 to @50 (36.39\% to 12.24\%) in Whoosh, amounting to 24.15\%, while in Solr, the drop is 45.58\%, yielding a variation margin of 21.43\%. Compared to the margin variation of 40.40\% for the SOTA method, the non-brute-force method, which relies solely on query content for query generation, demonstrates nearly twice the stability. In contrast, the SOTA brute-force technique, which utilizes both query and results, shows almost double the volatility. 

This behaviour (less volatility) can be observed in other NLP methods too as shown in Table \ref{tab:tfidf-evaluation} and \ref{tab:llm-evaluation}. Using `OR' between bi-grams and uni-grams leads to better scores, as shown in Table \ref{tab:tfidf-evaluation}. The Tf-Idf method yields lower scores compared to the YAKE method, as illustrated in Table \ref{tab:yake-evaluation}. This supports the conclusion that it is indeed challenging to retrieve patents or nearest neighbors based on embeddings derived from the Google Patents Research Data in BigQuery, provided by IFI CLAIMS Patent Services and Google. This observation suggests that embedding-based methods do not rely solely on the individual words but rather on the semantic meaning they convey. However, when attempting to predict using only words, they may not form a representative set. This pattern is consistently observed across all the NLP methods explored in this study. It highlights that embedding-based approaches and keyword-based searches are unlikely to share the same representative word clouds. 
\begin{table}
\centering
\caption{Evaluation Results of Tf-Idf\_Whoosh (\textbf{TW}) and Tf-Idf\_Solr (\textbf{TS}) (in percentages), \textbf{only query text} is used for query generation. \\
TW1: Without CPC, ADJ=1, AND used between bi-grams and uni-grams \\
TW2: With CPC, ADJ=1, OR used between bi-grams and uni-grams 
\\
TS1: Without CPC, proximity=1, AND used between bi-grams and uni-grams
\\
TS2: With CPC, proximity=1, OR used between bi-grams and uni-grams}
\begin{tabular}{llcccccc}
\toprule
\textbf{Method}   & \textbf{Metric}   & \textbf{@5} & \textbf{@10} & \textbf{@20} & \textbf{@30} & \textbf{@40} & \textbf{@50} \\
\midrule
\multirow{3}{*}{TW1} 
& MAP & 11.11 & 8.24 & 5.81 & 4.60 & 3.83 & 3.27 \\ 
    & Precision & 14.98 & 12.83 & 10.47 & 9.10 & 8.04 & 7.21 \\ 
    & Recall  & 1.49 & 2.56 & 4.19 & 5.46 & 6.43 & 7.21 \\\cline{2-8} 
\multirow{3}{*}{TW2} 
    & MAP & 35.95 & 28.09 & 19.77 & 15.15 & 12.11 & \textbf{10.03} \\ 
    & Precision & 42.56 & 35.62 & 26.72 & 21.25 & 17.42 & \textbf{14.70} \\ 
    & Recall  & 4.25 & 7.12 & 10.69 & 12.75 & 13.94 & \textbf{14.70} \\
\bottomrule
\multirow{3}{*}{TS1} 
& MAP & 28.76 & 21.12 & 10.56 & 7.04 & 5.28 & 4.22 \\ 
    & Precision & 33.19 & 26.34 & 13.17 & 8.78 & 6.58 & 5.26 \\ 
    & Recall  & 3.31 & 5.26 & 5.26 & 5.26 & 5.26 & 5.26 \\\cline{2-8} 
\multirow{3}{*}{TS2} 
    & MAP & 43.15 & 34.28 & 17.14 & 11.42 & 8.57 & 6.85 \\ 
    & Precision & 49.29 & 42.09 & 21.04 & 14.03 & 10.52 & 8.41 \\ 
    & Recall  & 4.92 & 8.41 & 8.41 & 8.41 & 8.41 & 8.41 \\
\bottomrule
\end{tabular}

\label{tab:tfidf-evaluation}
\end{table}

We employed the Mistral7B\footnote{\url{https://docs.mistral.ai/getting-started/models/models_overview/}} model, which has the capability to handle 32k tokens, to extract key terms from patent claims and descriptions. The corresponding retrieval results for the Mistral7B model are presented in Table \ref{tab:llm-evaluation}. The model demonstrates competitive performance compared to the Tf-Idf method. Additionally, there can be potential for further score improvements through advanced prompt engineering.

\begin{table}
\centering
\caption{Evaluation Results of LLM(Mistral7B)\_Whoosh (\textbf{LW}) and LLM(Mistral7B)\_Solr (\textbf{LS}) (in percentages), \textbf{only query text} is used for query generation. \\
LW and LS: With CPC, ADJ/proximity=1, OR used between bi-grams and uni-grams}
\begin{tabular}{llcccccc}
\toprule
\textbf{Method}   & \textbf{Metric}   & \textbf{@5} & \textbf{@10} & \textbf{@20} & \textbf{@30} & \textbf{@40} & \textbf{@50} \\
\midrule
\multirow{3}{*}{LW} 
& MAP & 31.08 & 24.65 & 17.67 & 13.75 & 11.11 & \textbf{9.26} \\ 
    & Precision & 37.71 & 32.59 & 25.23 & 20.60 & 17.16 & \textbf{14.64} \\ 
    & Recall  & 3.77 & 6.51 & 10.09 & 12.36 & 13.72 & \textbf{14.64} 
    \\\cline{2-8}
\multirow{3}{*}{LS} 
    & MAP & 41.36 & 32.98 & 16.49 & 10.99 & 8.24 & 6.59 \\ 
    & Precision & 47.79 & 40.71 & 20.35 & 13.57 & 10.17 & 8.14 \\ 
    & Recall  & 4.79 & 8.14 & 8.14 & 8.14 & 8.14 & 8.14 \\
\bottomrule
\end{tabular}

\label{tab:llm-evaluation}
\end{table}

In summary, the NLP methods explored in this work exhibit less volatility compared to the state-of-the-art (SOTA) techniques. While these methods may be less competitive relative to SOTA, several factors contribute to this, including the observations discussed in Section 3.2.1. Brute-force methods, particularly those that rely on both queries and results being known in advance, or that exploit statistical scores across databases prior to query generation, are not realistic for patent search applications.

This observation reinforces the central objective of this paper: there are significant challenges in benchmarking datasets and methods for patent prior art search and explainable AI. These challenges underscore the importance of adhering to rigorous evaluation practices and preserving traditional benchmark design principles, even as emerging technologies offer new possibilities.

Benchmark datasets play a critical role in evaluating the performance of embedding models for information retrieval tasks. The USPTO provides a query-nearest neighbor (NN) dataset for patent retrieval, where each query patent is associated with a set of 50 nearest neighbors, determined by an embedding model used by USPTO. While this dataset can be useful for evaluating specific retrieval methods, its design and underlying assumptions pose significant challenges for comparing different retrieval systems, particularly those based on dense embeddings. Therefore, in the following subsection, we delve into the investigation of using this dataset for retrieval by another dense vector embedding model.

\subsection{Limitations of Using the USPTO Query-Nearest Neighbor Dataset for Patent Retrieval Evaluation}
The core issue arises from the subjective nature of the nearest neighbor selection process, which is tightly coupled to the specific embedding model employed by the USPTO. The nearest neighbors (NNs) generated by one embedding model reflect the similarity criteria defined by that model, meaning other embedding models may produce equally relevant but different sets of NNs. This subjectivity introduces an inherent bias that complicates direct comparisons between models. To prove this hypothesis, we used a benchmarked embedding model from Huggingface, i.e., GTE Large\footnote{\url{https://huggingface.co/Alibaba-NLP/gte-large-en-v1.5 }}. In the Massive Text Embedding Benchmark (MTEB), this model is well known for its top performance in various NLP tasks, despite its smaller memory usage and larger context size. Therefore, we chose this model for evaluating the USPTO NN dataset. For a given query (out of a total of 3,343 test samples), we used the abstract, first claims, and summary created using extractive information within patents to generate embeddings. We used FAISS\footnote{\url{https://faiss.ai/index.html}} dense vector databases to perform the retrieval, and the results are recorded in Table \ref{tab:gte-large-evaluation}. 

\begin{table}
\centering
\caption{Evaluation Results of GTE-large model on USPTO NN dataset}
\begin{tabular}{llcccccc}
\toprule
\textbf{Method}   & \textbf{Metric}   & \textbf{@5} & \textbf{@10} & \textbf{@20} & \textbf{@30} & \textbf{@40} & \textbf{@50} \\
\midrule
\multirow{3}{*}{GTE-Abstract} 
& MAP & 44.26 & 35.92 & 27.47 & 22-63 & 19.99 & 16.60 \\ 
    & Precision & 51.18 & 44.76 & 37.41 & 32.83 & 29.31 & 26.49 \\ 
    & Recall  & 5.11 & 8.95 & 14.96 & 19.70 & 23.45 & 26.49
    \\\cline{2-8}
\multirow{3}{*}{GTE-1st\_Claim} 
    & MAP & 50.43 & 34.92 & 26.61 & 21.72 & 18.37 & 15.86 \\ 
    & Precision & 50.05 & 43.73 & 36.55 & 31.74 & 28.24 & 25.45 \\ 
    & Recall  & 5.00 & 8.74 & 14.62 & 19.04 & 22.59 & 25.45 
    \\\cline{2-8}
\multirow{3}{*}{GTE-Summary} 
    & MAP & 53.83 & 44.77 & 34.85 & 28.80 & 24.40 & \textbf{21.07} \\ 
    & Precision & 60.33 & 53.51 & 45.21 & 39.51 & 35.01 & \textbf{31.47} \\ 
    & Recall  & 6.03 & 10.70 & 18.08 & 23.70 & 28.00 & \textbf{31.47} \\
\bottomrule
\end{tabular}

\label{tab:gte-large-evaluation}
\end{table}

As we note in Table \ref{tab:gte-large-evaluation}, the results show that utilizing summaries instead of abstracts or claims has a greater impact on performance improvement in patent retrieval. Despite this, the scores still align closely with those generated by boolean queries, with only slight improvement. This raises several concerns, as outlined below:
\begin{itemize}
    \item \textbf{Bias Toward USPTO’s Embedding Model:} The 50 nearest neighbors (NNs) provided for each query are specific to the embedding model used by USPTO. Evaluating any model by comparing its retrieval results with this dataset essentially asks the model to replicate the behavior of USPTO’s embedding model. This poses a challenge, as different embedding models capture different semantic nuances or focus on different features (e.g., syntax, domain-specific terms, etc.). It’s possible that a model retrieves relevant patents that are not in the top 50 NNs identified by USPTO’s model but are still highly relevant. This implies that using the USPTO benchmark may unfairly penalize a model if it retrieves relevant patents that don’t match the NNs retrieved by USPTO’s model.
    \item \textbf{Subjectivity of Nearest Neighbors:} The concept of ``nearest neighbor" in embedding space is model-specific. A dense retriever optimized for different types of patent similarity (e.g., technical similarity, invention novelty) might retrieve a different set of neighbors than the ones listed in the USPTO dataset. For e.g. if our embedding model focuses more on semantic meaning in claims while USPTO’s model prioritized abstract-level similarity, our model might retrieve a different set of patents that are still highly relevant but not in the top 50 NNs.
    \item \textbf{No Ground Truth for Patent Similarity:} The NNs in the USPTO dataset are not necessarily the ground truth for patent similarity. They reflect the similarity as perceived by one model, but other models (or even human experts) may consider different patents to be more relevant. This creates an evaluation problem: if the 50 NNs aren’t a definitive benchmark for relevance, one is testing their model against a subjective standard that is only meaningful when replicating or comparing to that specific embedding model.
\end{itemize}
To further investigate this issue, we conducted an experiment comparing the USPTO dataset's nearest neighbors with the results produced by GTE-large dense model. We selected a sample query patent (US4279078A) and its extreme neighbors, i.e. 1st, 2nd, 49th, and 50th nearest neighbors from the USPTO dataset. Using GTE-large embedding model, we computed the cosine similarity between the query patent and these nearest neighbors, focusing on the first claim of the query patent and the description of the neighbors which are retrieved by USPTO's embedding model. Additionally, we calculated the similarity scores between the same query patent and the nearest neighbors retrieved based on the predictions of GTE-large model. The results are summarized in Table \ref{table:similarity_comparison} below.
\begin{table}
\centering
\caption{Comparison of Patent Similarity (\%) between USPTO NNs and GTE-large Model's predictions. \\
QC\_Top1\_D: query\_1st\_claim vs top1\_description \\ 
QC\_Top2\_D: query\_1st\_claim vs top2\_description \\ 
QC\_Top49\_D: query\_1st\_claim vs top49\_description \\ 
QC\_Top50\_D: query\_1st\_claim vs top50\_description 
}
\small 
\begin{tabular}{lcc}
\toprule
Patent Pair                          & USPTO NNs Similarity  & GTE-large NNs Similarity  \\ \hline
QC\_Top1\_D        & 75.83                             & 79.71                             \\ \midrule
QC\_Top2\_D        & 67.50                             & 78.88                             \\ \hline
QC\_Top49\_D       & \textbf{62.12 }                           & 71.83                             \\ \hline
QC\_Top50\_D       & 64.05                             & 71.67                             \\ \bottomrule
\end{tabular}

\label{table:similarity_comparison}
\end{table}
As seen in Table \ref{table:similarity_comparison}, there is a significant disparity between the similarities produced by the USPTO dataset's embedding model and those produced by GTE-large dense model. Notably, the 50th nearest neighbor in the USPTO dataset shows a higher similarity (64.05\%) with the query patent compared to the 49th nearest neighbor (62.12\%, in bold). This inconsistency suggests that the relative ranking of nearest neighbors in the USPTO dataset is not strictly determined by their semantic similarity to the query patent according to GTE-large. This discrepancy highlights a critical issue with using the USPTO dataset for evaluating retrieval models: \emph{the nearest neighbors provided are inherently tied to the specific embedding space of the USPTO's model and may not reflect the actual relevance of the retrieved patents in other embedding spaces.}

\subsection{Considerations for Future Evaluation and Benchmark Comparison}
\label{sec:FutureDesign}

To advance research in patent text mining, it is essential to address several critical issues highlighted by recent findings in the broader field of large language models (LLMs). Studies in computational linguistics have revealed substantial concerns regarding data contamination and evaluation practices. For example, a systematic review of 255 studies on OpenAI's GPT-3.5 and GPT-4 found that these models had indirect exposure to approximately 4.7 million samples from 263 different datasets. Such exposure introduces potential biases in evaluation outcomes, as models may be inadvertently tested on data they have previously encountered.

The review also identified prevalent evaluation malpractices, including inadequate baseline comparisons, inconsistent evaluation scales, and limited reproducibility and transparency. Many studies either fail to benchmark LLMs against alternative methods or provide restricted access to data and code, undermining the reliability and fairness of reported results. Similar challenges are emerging in the patent text mining community, as demonstrated in our analyses of USPTO and MoZIP benchmarks, highlighting the need for more rigorous evaluation standards.

To address these challenges, we propose the adoption of improved evaluation protocols and greater transparency when working with closed-source models. Best practices include detailed documentation of experimental methodologies, strict adherence to data usage policies, and the development of robust frameworks for assessing LLM performance. Such measures ensure that evaluations are fair, reproducible, and representative of real-world patent search scenarios.

We recommend organizing evaluation criteria along three dimensions:

\textbf{Experiment Details:}
\begin{enumerate}
    \item \textit{Applicability:} Specify whether the experiment evaluates third-party LLMs, trains models from scratch, or fine-tunes existing models.
    \item \textit{Date of Experiment:} Record the earliest known draft or version of the experiment.
    \item \textit{Running Period:} Indicate the dates during which the experiment was conducted.
\end{enumerate}

\textbf{Methodology and Transparency:}
\begin{enumerate}
    \item \textit{Prompt Availability:} Indicate whether experiment prompts are provided.
    \item \textit{Code Repository:} Note if the code is publicly accessible.
    \item \textit{Model Identification:} Specify which model(s) were used.
    \item \textit{Model Access:} Clarify whether the model was accessed via API or interface.
    \item \textit{Sampling Methods:} Describe how sampling prevents exact data reconstruction.
    \item \textit{Baseline Comparisons:} Indicate whether other methods were included for benchmarking.
    \item \textit{Evaluation Scale Consistency:} Note discrepancies between scales for different models.
\end{enumerate}

\textbf{Evaluation and Data Handling:}
\begin{enumerate}
    \item \textit{Benchmark Usage:} Specify if known benchmark datasets were adapted, and whether evaluation metrics follow benchmark conventions.
    \item \textit{Custom Datasets:} Indicate whether ad-hoc datasets were used and their availability.
    \item \textit{Data Leakage:} Report any direct or indirect exposure of test or training data, including dataset name, task, split, and fraction of data affected.
\end{enumerate}

Here, \textit{data leakage} refers to unintended access to sensitive or proprietary data during training or evaluation. We distinguish:

\begin{enumerate}
    \item \textit{Indirect Leakage:} Occurs when external datasets or benchmarks are inadvertently exposed to the model, potentially biasing evaluations.
    \item \textit{Direct Leakage:} Occurs when training, validation, or test data is explicitly included in model training, leading to overfitting and inflated performance metrics.
\end{enumerate}

Data leakage undermines the validity and integrity of experimental results, particularly when proprietary or sensitive information is involved. Addressing these issues is crucial for designing robust, reproducible, and fair evaluations in patent text mining and beyond.

Together, the analyses of MoZIP and the USPTO Kaggle challenge demonstrate how evaluation outcomes in patent
retrieval can be strongly shaped by benchmark construction, relevance definitions, and experimental constraints.
These findings motivate a broader reflection on how emerging AI-based retrieval systems should be assessed.
The concluding section synthesizes these insights and discusses their implications for future benchmark design and
evaluation practice in patent information retrieval.

\section{Conclusion}
This paper examined the challenges of evaluating information retrieval methods in the patent domain, with a particular focus on recent benchmarks designed to assess large language models. Through an in-depth investigation of the USPTO Explainable AI for Patent Professionals Kaggle challenge and a critical analysis of the MoZIP benchmark, we demonstrated that reported performance outcomes are often shaped more by benchmark design choices than by genuine advances in prior art search capability.
In the USPTO challenge, exceptionally high scores were achieved through strategies that exploit weaknesses in query constraints, evaluation metrics, and model-specific nearest-neighbor ground truth. Similarly, our analysis of MoZIP revealed that quiz-based task formulations, model-dependent relevance construction, and the absence of strong retrieval baselines limit its suitability for evaluating patent retrieval systems. In both cases, relevance is implicitly defined by the assumptions embedded in the benchmark itself, rather than by independent, task-appropriate judgments aligned with professional patent search practice.
These findings reinforce long-standing insights from patent information retrieval research: meaningful evaluation requires benchmarks that reflect the linguistic complexity of patent texts, the recall-oriented nature of prior art search, and the legal and technical criteria that define relevance. The CLEF-IP collections remain a valuable reference in this regard, demonstrating how robust benchmark design enables fair comparison across diverse retrieval approaches. As LLMs continue to evolve, progress in patent retrieval will depend not only on more powerful models, but also on the careful preservation and adaptation of established evaluation principles. Without such rigor, benchmarks risk measuring conformity to dataset artifacts rather than true advances in patent information retrieval.

\section*{Declaration on the Use of Generative AI}

The authors declare that generative artificial intelligence tools (GPT-5 and DeepSeek) were used solely for
language-related assistance, including grammar correction, spelling correction, and stylistic refinement of the
manuscript. These tools were not used to generate scientific content, interpret results, or draw conclusions.

The corresponding author takes full responsibility for the integrity, originality, and accuracy of the manuscript,
and for ensuring that the use of generative AI complies with the policies of the publisher and relevant ethical
guidelines.

\bibliographystyle{unsrtnat}
\bibliography{references}  






\end{document}